\documentclass[conference]{IEEEtran}
\IEEEoverridecommandlockouts

\usepackage{amsmath,amssymb,amsfonts}
\usepackage{mathtools}
\usepackage{algorithmic}
\usepackage{graphicx}
\usepackage{textcomp}
\usepackage{xcolor}
\usepackage{subcaption}
\usepackage{fixmetodonotes}
\usepackage[sorting=none]{biblatex}
\usepackage[acronym]{glossaries}
\usepackage{booktabs}

\usepackage{tikz}

\newcommand\copyrighttext{%
  \footnotesize \textcopyright 2024 IEEE. Personal use of this material is permitted.
  Permission from IEEE must be obtained for all other uses, in any current or future
  media, including reprinting/republishing this material for advertising or promotional
  purposes, creating new collective works, for resale or redistribution to servers or
  lists, or reuse of any copyrighted component of this work in other works.}
\newcommand\copyrightnotice{%
\begin{tikzpicture}[remember picture,overlay]
\node[anchor=south,yshift=10pt] at (current page.south) 
  {\fbox{\parbox{\dimexpr\textwidth-\fboxsep-\fboxrule\relax}{\copyrighttext}}};
\end{tikzpicture}%
}

\makeglossaries
\newacronym{mc}{MC}{Markov chain}
\newacronym{mae}{MAE}{mean absolute error}
\newacronym{mse}{MSE}{mean squared error}
\newacronym{rmse}{RMSE}{root mean squared error}
\newacronym{gmm}{GMM}{Gaussian Mixture Model}

\def\BibTeX{{\rm B\kern-.05em{\sc i\kern-.025em b}\kern-.08em
    T\kern-.1667em\lower.7ex\hbox{E}\kern-.125emX}}
\begin{document}

\title{Advancing Standard Load Profiles with Data-Driven Techniques and Recent Datasets \\
\thanks{This work has received funding from the European Commission under the Grant Agreement 101136119.}
}


\author{\IEEEauthorblockN{ Jawana~Gabrielski,  Ulf Häger}\\
\IEEEauthorblockA{\textit{Institute of Energy Systems, Energy Efficiency and Energy Economics}\\
TU Dortmund University, Dortmund, Germany}\\
jawana.gabrielski@tu-dortmund.de, ulf.haeger@tu-dortmund.de}

\maketitle
\copyrightnotice


\begin{abstract}
Estimating electricity consumption accurately is essential for the planning and operation of energy systems, as well as for billing processes. Standard Load Profiles (SLP) are widely used to estimate consumption patterns of different user groups. However, in Germany these SLP were formulated using historical data from over 20 years ago and have not been adjusted since. Changing electricity consumption behaviour, which leads to increasing deviations between load patterns and SLP, results in a need for a revision taking into account new data. The growing number of smart meters provides a large measurement database, which enables more accurate load modelling. This paper creates updated SLP using recent data. In addition, the assumptions of the SLP method are validated and improvements are proposed, taking into account the ease of applicability. Furthermore, a Fourier Series-based model is proposed as an alternative SLP model. The different models are compared and evaluated.
\end{abstract}


\section{Introduction}
The estimation of electricity consumption is a crucial aspect of the planning and operation of energy systems as well as billing processes~\cite{boogen2021estimating}. Standard Load Profiles (SLP) provide a simple approach and are widely used to estimate consumption patterns of different user groups. These SLP were formulated by the German Association of Energy and Water Industries (German: Bundesverband der Energie und Wasserwirtschaft, BDEW) using historical data from over 20 years ago~\cite{meier1999repraesentative} and have not been adjusted since. However, changing electricity consumption patterns due to changing behaviour as well as evolving devices, lead to increasing deviations between load patterns and SLP~\cite{bruckmeier2017teilbericht}. As a result, a revision based on new data is required. 
The paper focuses on residential SLP. In Germany, the residential sector accounts for 28~\% of total energy consumption~\cite{bdew_stromverbrauch}. The growing number of smart meters provides a larger measurement database, enabling more accurate load modelling. 
As energy systems transition toward greater reliance on decentralized and renewable energy sources, residential flexibility becomes increasingly important. Accurate and up-to-date SLP are a significant factor in enabling this flexibility by reflecting current consumption behaviors and supporting demand response initiatives.
In order to provide an insight into the existing SLP and other modelling approaches, these are first explained in section~\ref{sec2}.
As the data used for the conventional SLP is not publicly available, this paper first creates an updated SLP based on recent open data (section~\ref{sec3}). The resulting updated SLP is not only compared to the conventional SLP, but also serves as a benchmark for subsequent comparisons with alternative methods. This approach facilitates a direct and unbiased assessment of different modelling techniques, eliminating the interacting effects arising from different data sources. Secondly, the assumptions of the conventional BDEW SLP method are validated and improvements in the modelling method are proposed (section~\ref{sec4}). Finally, a Fourier Series model is developed as an alternative SLP modelling approach (section~\ref{sec5}). The different methods are compared and evaluated. Section~\ref{sec6} summarizes the findings and gives an outlook.
Other research in this area, which deals with the use of new data for building new SLP, mainly focusses on the examination of regional differences as in~\cite{scholz2017how} and~\cite{hinterstocker2014bewertung}, however, structural deviations are already evident there. Further research investigates the consideration of new typical days as Fridays, holidays, or bridge days, but concludes that the benefit of using new typical days is marginal~\cite{staats2018optimierung}. However, the advancement of the actual method is barely investigated. 

\section{Theoretical Background} \label{sec2}
The methods described in this paper are strongly related to the existing SLP method as well as other time series models. Hence, the relevant theoretical backgrounds are explained first. 

\subsection{Conventional Standard Load Profile}\label{subsec:SLP}
The conventional SLP model uses a multi-step modelling approach based on aggregated load data. In the first step, each individual consumption time series is scaled to 1000 kWh per year and the scaled time series are aggregated. The second step computes the yearly seasonality of the time series, fitting a polynomial to the annual course of daily energy consumption, which is referred to as the \textit{dynamisation factor}. In the third step, daily seasonalities are calculated per season. For this, three seasons are used: summer, winter, and transition, which summarizes spring and autumn. Before the daily seasonalities are calculated, each day of the yearly time series is divided by the corresponding dynamisation factor, so that the yearly seasonality is not accounted for twice. Based on this new annual time series, daily profiles are generated, calculating the quarter-hourly average for each season separately for workdays, Saturdays, and Sundays. National holidays are treated as Sundays. Finally, to obtain a yearly SLP, the daily profiles are multiplied by the dynamisation factor, so that discontinuities between the seasons are reduced. Since at the time when the SLP was developed measuring equipment was not yet so widespread, measurements were carried out over several weeks per season instead of using measurements for an entire year~\cite{meier1999repraesentative}.

\subsection{Time Series Modelling}\label{Fourier}
In recent years, much research has been conducted in the field of time series analysis and new models have been developed. Most time series models follow a decomposition approach to separate the underlying seasonalities. The two primary types of decomposition are additive and multiplicative. The first assumes that the different components accumulate linearly, whereas the second allows to consider multiplicative interactions~\cite{sinoquet2021time}. The different models furthermore distinguish in how they model the different components. The conventional SLP model can be described as a type of multiplicative model, which uses different kinds of averages. Another decomposition approach are Fourier Series, which are mathematical methods used to decompose periodic functions into a sum of sinusoidal waves. This technique can be applied to time series data to model seasonal patterns. General additive or multiplicative Fourier Series models can also combine the seasonalities, which are extracted using Fourier Series, with components like trend or additional repressors~\cite{taylor2017forecasting}.  

\section{Updated Standard Load Profile}\label{sec3}

In this section, a new SLP based on recent data is developed.
Summerizing the model explained in \ref{subsec:SLP} leads to \eqref{eq:slp}. 
\begin{equation}
\begin{split}
X(t) = d (t) & \cdot \sum_{i=1}^{3}\sum_{j=1}^{3} 
\bigl( c_i^{wd}(t) \cdot  c_{j}^s (t) \cdot s_{wd,s} (t) \bigl)  + \epsilon(t)
\end{split} 
\label{eq:slp}
\end{equation}
Where $X$ refers to the load value at time $t$, $d$ presents the dynamisation factor, $s_{wd,s}$ the daily seasonality of the considered weekday and season. The varying daily seasonalities per season and weekday are taken into account by means of the
binary variables $c_{i}^{wd}$ and $c_j^s$  which are one in case $t$ is within the considered weekday or season and zero if not. 
The error term $\epsilon$
represents changes which are not represented by
the model. 
This method is applied to data from the openMeter platform~\cite{logarithmo_openmeter} that provides anonymized smart meter measurements from more than 1000 residential consumers from Germany for several years. As the time series are real-world measured data, they may contain missing or defective data, which are identified as described in~\cite{gabrielski2023markov}; defective data is deleted. 
\begin{figure}[bth]
\centering
     \includegraphics[width=0.44\textwidth]{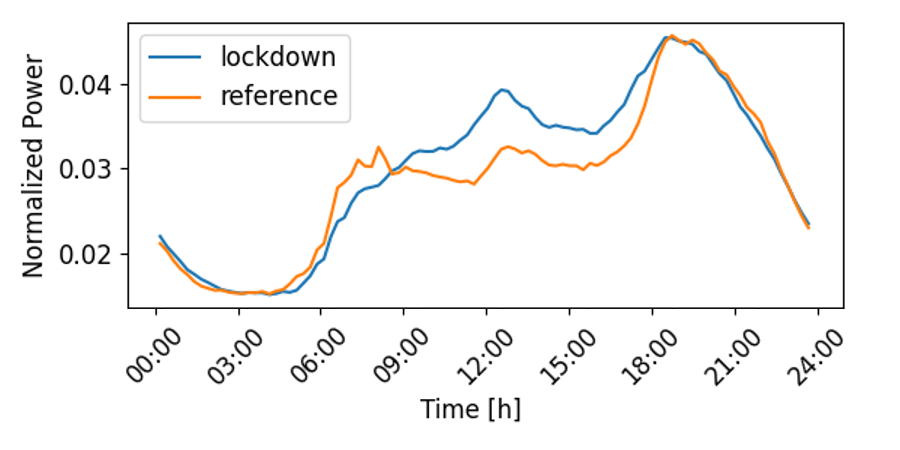}
      \caption{Comparison of weekday daily lockdown to not-lockdown profiles (scaled to an annual consumption of 1000~kWh)}
       \label{lockdown}
\end{figure}

To apply the SLP method all time series have to be scaled to a yearly consumption of 1000 kWh. Given the strong yearly seasonality, it is unfavourable to scale data with a large proportion of missing data. Hence, only data where at least 95~\% of the year is available are used and scaled percentually. The first measurements from the used source began in year 2017 and the latest are still being conducted to this day. Most of the data stems from the years 2020 and 2021. During this period, multiple lockdowns took place, due to the COVID-19 pandemic. Thus, it is first examined, whether the lockdowns influenced the residential load behaviour. For this, average daily profiles from before and after the lockdown times, are compared to average profiles during lockdown, taking into account the different types of weekdays as well as the time of the year. The results show a significant change in load behaviour, as depicted in Fig.~\ref{lockdown}. Hence, data from lockdown times was excluded from creating SLP. Most of the considered data from the openMeter platform stem from residential loads from North Rhine-Westphalia, Baden-Württemberg, Bavaria, Hesse and Lower Saxony with fewer measurements from new federal states as well as the North of Germany. However, the BDEW SLP also only considers measurements from former German states, making the presented approach consistent. 
As the provided time series are not labelled and include different consumers like households with e.g. photovoltaic power plants, or electric vehicles, some pre-processing is required first, in order to extract relevant consumers. In the course of this, time series with photovoltaic generation, as well as with electric vehicle consumption are excluded.
The remaining time series are used to apply the SLP method. To take into account the varying number of measurements during the considered time, the average value is weighted with the number of available measurements for each year corresponding to the parity of time as explained in~\cite{baranek2013optimierung}. Accordingly, the individual time series are weighted equally for each available timestep.

\begin{figure}[tbh]
\centering
     \includegraphics[width=0.48\textwidth]{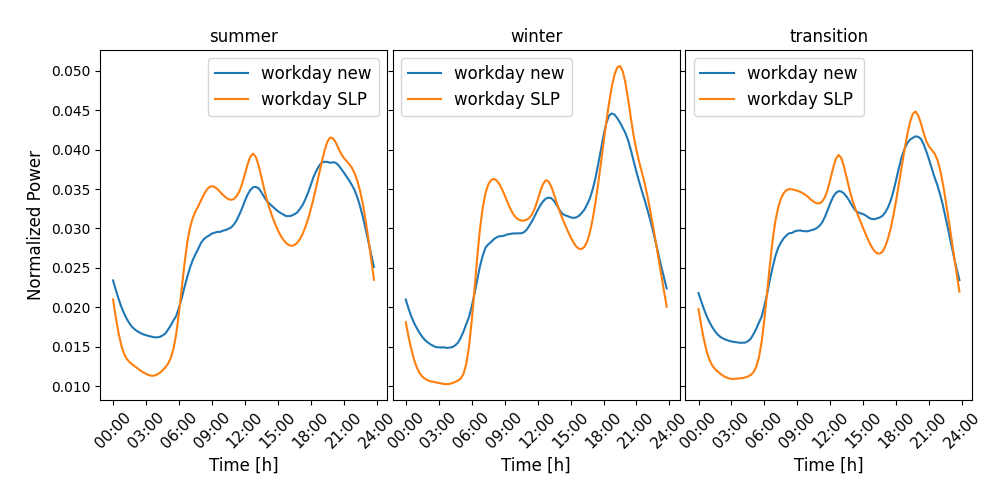}
      \caption{Comparison of daily workday seasonalities between SLP and new data (scaled to an annual consumption of 1000~kWh)}
       \label{comp_SLP_new}
\end{figure}

Fig.~\ref{comp_SLP_new} compares the obtained daily profiles for workdays of the different seasons. The general course is still similar, however, three structural differences are noticeable: The morning peak is less significant, which could be explained by more flexible working hours as well as a lower penetration of electrical water heaters, which are used to heat shower water in the morning. Furthermore, smart home technologies like time-controlled dishwashers or washing machines might shift part of the morning peak to later times of the day. Moreover, the day-night seasonality is lower, which could be a consequence of more stand-by appliances or in general higher energy efficiency~\cite{odyssee_mure}.  
Additionally, the midday peak is reduced, which can be attributed to changing lifestyles, as more women are now part of the workforce~\cite{pfahl2024erwerbstaetigenquoten} no longer staying at home to cook and care for their children during the day. Besides, stoves with higher efficiency~\cite{odyssee_mure} as well as increased food delivery might contribute to reduced midday peaks.

\begin{figure}[thb]
\centering
     \includegraphics[width=0.45\textwidth]{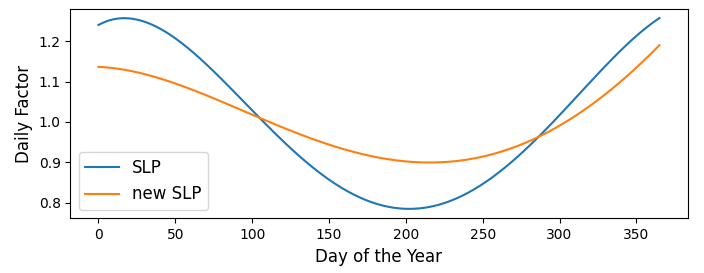}
      \caption{Comparison of dynamisation factor from conventional SLP and new data}
       \label{dyn}
\end{figure}

Fig.~\ref{dyn} compares of the dynamisation factor for the conventional and the updated SLP. It is remarkable, that the new SLP has lower yearly seasonality, which could be a result of more efficient appliances like LED light bulbs which consume less energy, and fewer temperature-dependent consumers like electric water heaters.

\begin{figure}[bht]
\centering
     \includegraphics[width=0.48\textwidth]{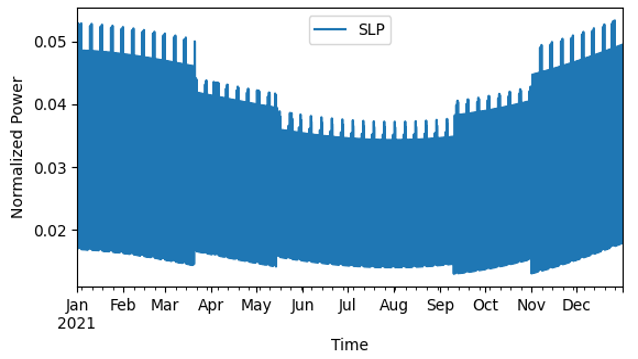}
      \caption{SLP based on new data (scaled to an annual consumption of 1000~kWh)}
       \label{BDEW_SLP}
\end{figure}

Fig.~\ref{BDEW_SLP} shows the resulting SLP based on new data. It should be considered that the data were collected around the time of the COVID-19 pandemic, which might result in higher shares of people being at home and working from home. In addition, it is noteworthy, that the smart meter data collected in the openMeter project comes from people, who actively decided to install and pay for a smart meter, hence this group might not correspond to the socio-economic average. For example, the reduced morning peak could indicate, that many consumers work in home office. However, as the paper mainly focusses on the method to build the SLP, the resulting new SLP is used as a comparison for the following analysis.

\section{Validation and Enhancement of the Standard Load Profile method}\label{sec4}
In this section the existing SLP is validated leveraging data-drive- techniques and on this basis improvements are proposed. 
\subsection{Validation of weekday assumption}
The conventional method is, as described in section~\ref{subsec:SLP}, based on the assumption that weekdays can be differentiated between workday, Saturday, and Sunday. National holidays are considered as Sundays and Christmas Eve as well as New Year’s Eve as Saturday. To validate this, different machine learning time series classification models are trained to distinguish between workday, Saturday, and Sunday. The best results are obtained with a Random Forest classification, which can predict the day type with a mean accuracy of 0.96. This model is applied to classify national holidays; the majority of them are categorized as Sundays.
Christmas Eve, which is no national holiday, is classified as Saturday (besides being Sunday). New Year’s Eve, which is also no national holiday, is categorized as Sunday. So it is validated that national holidays are considered as Sunday and Christmas Eve as Saturday. As opposed to the conventional method, New Year’s Eve is considered as Sunday. 
\subsection{Validation of season assumption}
\begin{figure}[thb]
\centering
     \includegraphics[width=0.35\textwidth]{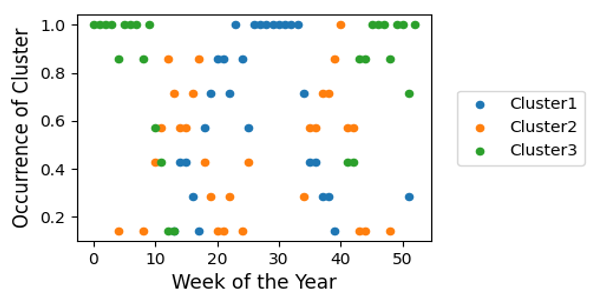}
      \caption{Normalized cluster assignment per week of the year}
       \label{clust_affil}
\end{figure}

The conventional SLP method defines three seasons. To validate this assumption, a time series k-means clustering approach is applied to the average daily profile of each day of the year. In order to take into account only the shape of the profile, a min-max scaling transformation is used. The optimal number of clusters is chosen in accordance with the silhouette score, which takes into account how well data fits into the assigned cluster and how it differs from others~\cite{rousseeuw1987silhouettes}. Based on this, three clusters are chosen. 
Fig.~\ref{clust_affil} shows their occurrence per week. 
Analysing the distribution of the assigned clusters during the year shows the same three seasons used in the conventional method, dividing the daily profiles into summer, winter, and transition.
\subsection{Enhancements}
The assignment can furthermore be used to identify the transition point between the seasons, which differs from those of the conventional SLP. 
Moreover, it can be observed, that there is no distinct transition point, but rather a sliding transition with partially alternating assignments.
In the conventional model the multiplication of the yearly polynomial with daily seasonalities per season leads to abrupt changeovers between the different seasons (as can be seen in Fig.~\ref{BDEW_SLP}), which do not reflect these sliding transition. To take into account the transition, a linear combination with time-varying weights is implemented for the changeovers, as shown in \eqref{y_alpha}.
\begin{equation}
y(\alpha) = (1 - \alpha) \cdot p_{s1} + \alpha \cdot p_{s2}
\label{y_alpha}
\end{equation}
$p_{s1}$ and $p_{s2}$ are the daily profiles of the different seasons and $y$ is the resulting profile during the transition. $\alpha$ is the normalized distance to the transition time and can be calculated taking into account the transition time $t_t$ and the transition duration $d_t$ as described in \eqref{alpha}.
\begin{equation}
\alpha = \Biggl|\frac{t_t - d_t}{d_t}\Biggl|
\label{alpha}
\end{equation}
This method implies the assumption, that the transition between the seasons is formed from a mixture of both adjacent seasons; the smaller the distance to the adjacent season, the greater the share of this season in the resulting profile. The underlying assumption can be validated by applying k-means with four or more clusters, which results in more granular transition clusters.

\begin{figure}[hbt]
\centering
     \includegraphics[width=0.45\textwidth]{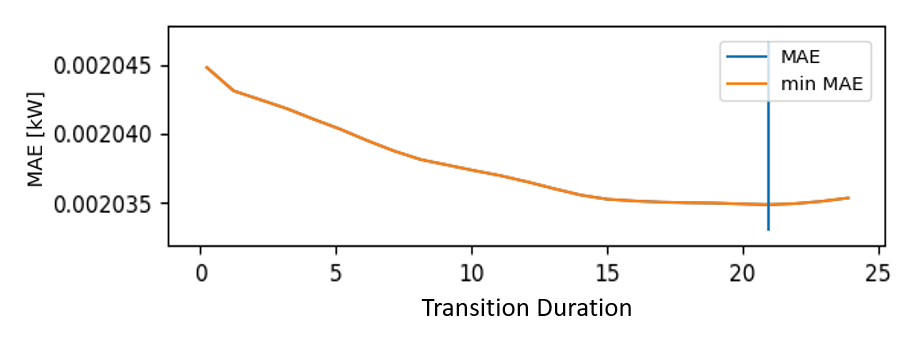}
      \caption{Error for different transition durations}
       \label{min_mae}
\end{figure}
\begin{figure}[bth]
\centering
     \includegraphics[width=0.48\textwidth]{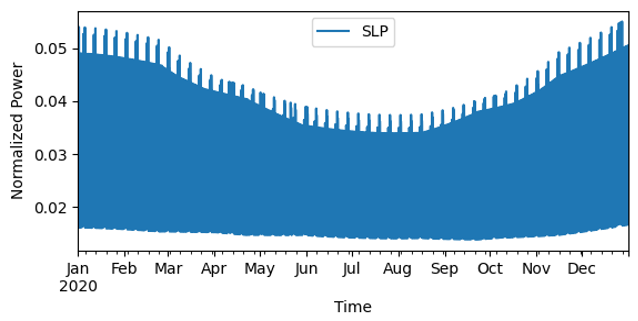}
      \caption{Resulting new SLP (scaled to an annual consumption of 1000~kWh)}
       \label{newslp}
\end{figure}

Different durations of transition are tested and the error is calculated, using the mean absolute error (MAE) between the average time series and the profile, which are both scaled to 1000~kWh per year. The error is shown in Fig.~\ref{min_mae}. The minimal error occurs with a transition duration of 21 days, which is hence proposed for the new SLP method. Fig.~\ref{newslp} shows the resulting yearly SLP. Compared to Fig~\ref{BDEW_SLP}, there are no longer abrupt transitions between seasons.

To evaluate the proposed method, the error between the average input time series and the model is calculated. This approach is used due to the limited amount of data being available, not allowing to fit models with training and validation data, as this would lead to higher variability and hence less accuracy. Table~\ref{tab:error} shows the MAE for the different models.

\begin{table}[tbh]
\begin{tabular}{llll}
    \toprule
    & \begin{tabular}[c]{@{}l@{}}BDEW  with\\ new data\end{tabular} & \begin{tabular}[c]{@{}l@{}}With adapted\\  weekdays and  seasons\end{tabular} & \begin{tabular}[c]{@{}l@{}}With linear combination\\ for transition duration\end{tabular}   \\
     \midrule
MAE & 0.002048                                                        & 0.002045                                                                        & 0.002035      \\
\bottomrule
\end{tabular}
\caption{Evolution of MAE (in kW) for different models}
\label{tab:error}
\end{table}

In addition to the multiplicative approach, the model is tested as an additive model. The additive approach leads to higher errors, as it is not capable of modelling the influence of the yearly seasonality on the daily seasonality. 

Despite more than 1000 yearly time series being used to generate the SLP, in particular, the weekend daily profiles (Saturday and Sunday) retain some of the original variability and are not fully smoothed. Hence, it is proposed to apply a filter to the daily profiles. The Savitzky- Golay filter~\cite{savitzky1964smoothing} is a low pass filter that fits a polynomial to a window of adjacent points and uses it to estimate the smoothed value in the middle of the window. It is capable of smoothing noisy data while preserving the shape and height of peaks in the data. 

\begin{figure}[bth]
\centering
     \includegraphics[width=0.48\textwidth]{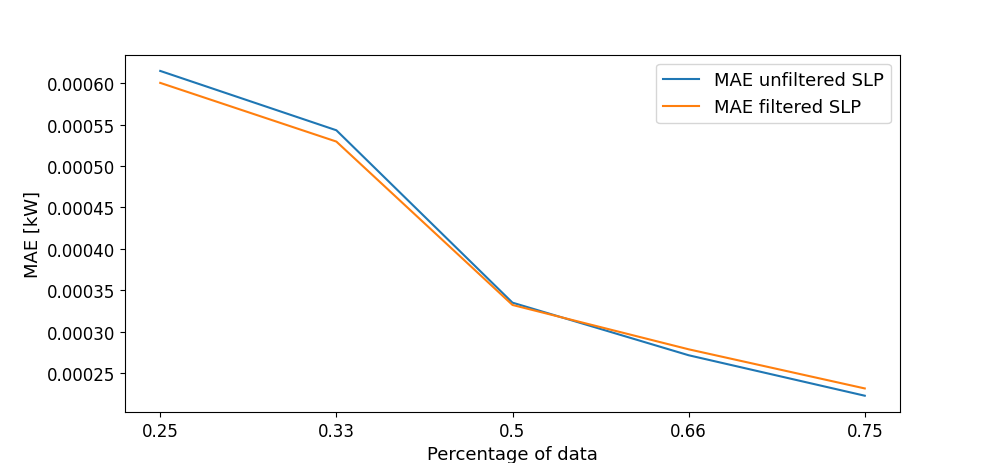}
      \caption{Error between final profile (without filter) and profile of different shares of data, for filtered and unfiltered data respectively}
       \label{err_comp}
\end{figure}

The challenge in determining whether the filter reduces the error is that there is not sufficient data to actually test it, as using only half of the data does not lead to accurate profiles. Hence, in order to test if the filter is able to model the data more accurately, different shares of data are used to build the SLP. The profiles, made from smaller shares of data, are generated with and without applying the filter; the error compared to the whole profile shown in Fig.~\ref{newslp} is compared for filtered and unfiltered data. As the error highly depends on the used data, the profiles from smaller shares of measurements are built ten times for randomly selected time series, and the error is averaged. Fig.~\ref{err_comp}  shows these average errors between the resulting profile and the different filtered and unfiltered profiles generated from different  shares of data. It becomes clear, that the error of the filtered profile is lower than the error of the unfiltered profile as long as not much more than half the data is used. Clearly, using more than half of the data generates a profile, which closely aligns with the one generated based on the whole data. Hence, only the first half of the data is used to evaluate the filter. As the error of the average profile is lower in this part, it can be inferred, that the filter reduces the error and thus, can be used to estimate a profile, generated based on more data. 

\begin{figure}[bht]
\centering
     \includegraphics[width=0.48\textwidth]{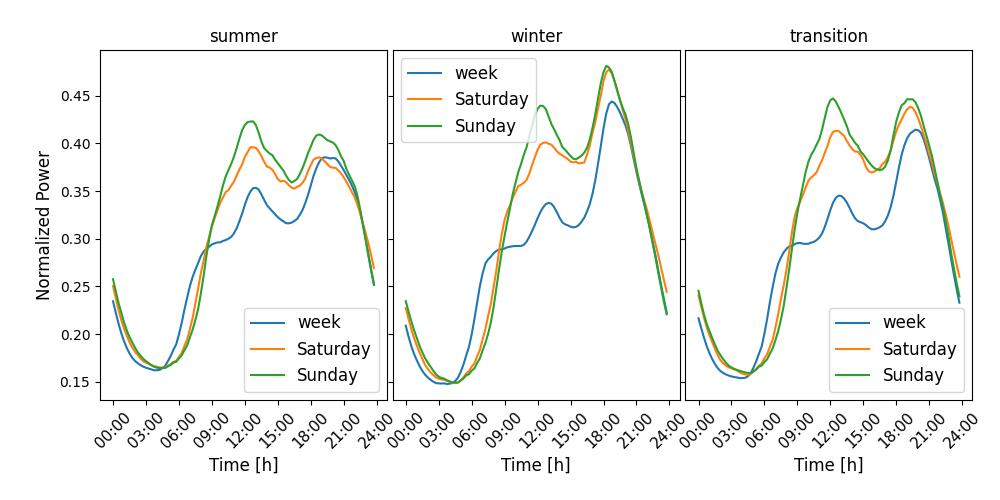}
      \caption{Resulting daily profiles (scaled to an annual consumption of 1000~kWh)}
       \label{d_prof}
\end{figure}

Additionally,  Fig.~\ref{err_comp} could be used to determine which minimum number of time series in required to form a SLP. 
At 33~\% of the data being used,  which are approximately 330 time series, a kink can be observed. 
It could be concluded, that at least 300 time series should be used. Another kink is visible at 50~\% of the data, which are approximately 500 time series. That could indicate that at this number the quality of the SLP increases further. These assumptions also roughly match the minimum requirement of \textit{a few hundred} mentioned in~\cite{meier1999repraesentative}. However, they should be validated with more data. Fig.~\ref{d_prof}  shows the resulting filtered daily profiles. 

\begin{figure}[bth]
\centering
     \includegraphics[width=0.5\textwidth]{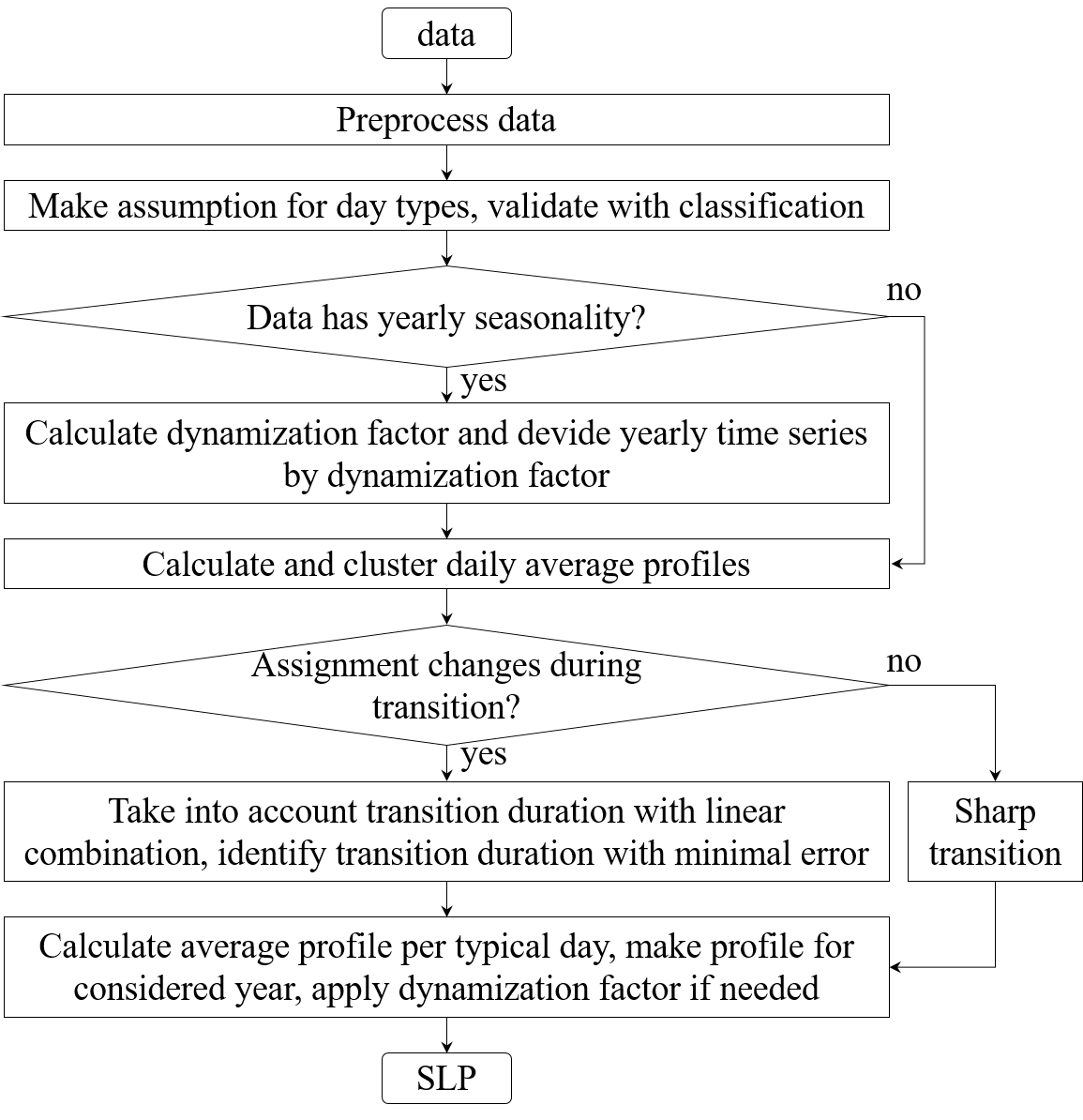}
      \caption{Flow chart of the proposed SLP method}
       \label{flowchart}
\end{figure}

The proposed method is, in fact, independent of the data used. Without any prior assumptions being made, it could be applied to model not only household electricity consumption but also other consumption or in general periodic time series. The method is depicted in the flow chart in Fig.~\ref{flowchart}. 

\section{Comparison to Fourier based Model}\label{sec5}
In addition to the improvement of the conventional SLP, a Fourier-series-based approach is presented. 
The general Fourier approach described in section~\ref{Fourier} can be adapted to take into account different seasonalities for workday and weekend as presented in \eqref{eq:four}. 
\begin{equation}
\begin{split}
X(t) = s_y (t) & + \sum_{i=1}^{3} 
\bigl( c_i^s(t) \cdot  c_{wd} (t) \cdot s_{wd,s} (t) \\\
&+  c_i^s \cdot(1 - c_{wd} (t)) \cdot s_{we,s} (t)\bigl)  + \epsilon(t)
\end{split} 
\label{eq:four}
\end{equation}

Where $X$ refers to the load value at time $t$, $s_y$ presents the yearly seasonality, $s_{wd,s}$ the daily workday seasonality of the considered season, and $s_{we,s}$ the daily weekend seasonality. The error term $\epsilon$
represents changes which are not represented by
the model. 

 \begin{equation}
    c_{wd} (t)= 
\begin{dcases}
    1,& \text{if  } weekday(t)<6\\
    0,          & \text{if } weekday(t)\geq6
    \label{eq:cases}
\end{dcases}
 \end{equation}
The different daily seasonalities are considered by means of a binary variable $c_{wd}$ which is one in case $t$ is a workday and zero in case $t$ is weekend, as shown in \eqref{eq:cases}. $c_i^s$ are also binary variables indicating the current season. Moreover, it is possible to add further seasonalities, e.g., weekly seasonalities or individual seasonalities for Saturday and Sunday. In this case, the different seasonalities are summed up as well to form the overall model. Modelling seasonalities with Fourier Series ensures a continuous transition between the beginning and the end of the period and hence both are at the same level. 
A Fourier Series-based model has the advantage that different seasonalities can be combined easily. Hence, another seasonality, which considers the weekly seasonality, is added to the model as well as three different daily seasonalities for workday, Saturday, and Sunday as in the SLP. 
The described approach is tested with additive and multiplicative models. The additive model has better results and is used in the following. 
Comparing the daily seasonalities of the SLP and the Fourier model, it is noticeable, that the profiles are quite similar, however the Fourier approach leads to more smoothed profiles. 

\begin{figure}[tbh]
\centering
     \includegraphics[width=0.48\textwidth]{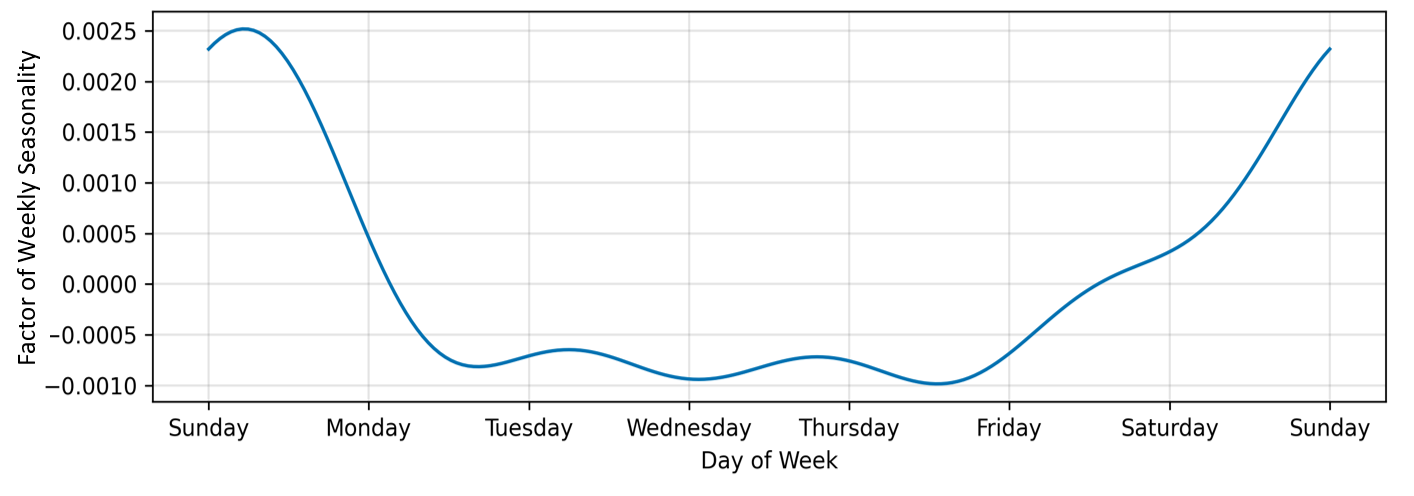}
      \caption{Additive factor of weekly seasonality}
       \label{weekly}
\end{figure}

Fig.~\ref{weekly}  illustrates the weekly seasonality. It highlights a noticeable increase during the weekend, beginning on Friday afternoon. While higher values during the weekend can be considered by the daily profiles, the elevated consumption on Friday afternoon cannot be modelled by the uniform workday profile and hence is not taken into account by the SLP method. 

The resulting profile which looks quite similar to the SLP, is not shown here. The error is slightly higher compared to the conventional SLP (the MAE is 0,002117~kW per timestep for a yearly consumption of 1000~kWh). Analysing the error in the course of the year reveals that the error is especially higher at the beginning and the end of the year. That can be explained by the fact that Fourier Series aim at having the same level at the beginning and the end of each considered period. This is not appropriate for the load profile, as the load at the end of the year has a high increase due to national holidays. 
In total, it can be summarized, that the SLP method is more suitable for residential consumption patterns.

\section{Conclusion and Outlook}\label{sec6}

In summary, it is demonstrated that load behaviour has changed and BDEW SLP are no longer up to date. In addition, the assumptions of the conventional SLP method are scientifically validated and modified, leading to an enhancement, still taking into account the ease of applicability. Furthermore, a general additive Fourier model is proposed and compared to the SLP. It is observed that the Fourier model is less suitable to model residential loads. 
However, its ability to combine different seasonalities enables it to model varying daily seasonalities in the course of the workweek, e.g. on Friday.
This could be considered in future SLP as well. 

Further improvements could be made with larger datasets. 
Despite, the lockdown periods being excluded from the data used for this paper, the data was measured around the COVID-19 pandemic and behavioural changes due to the pandemic might influence the resulting profile. Additional research should investigate the change in behaviour and consumption, before, during, and after the pandemic. Moreover, it should be tested, whether the proposed filtering approach is appropriate for larger datasets. Finally, the accuracy of the enhanced method should be quantified based on new data. 
Additionally, the integration of advanced data management systems, like digital twin technologies and real-time data analytics could provide improvements in future electricity consumption models by utilizing large amounts of data as well as real-time data.

\printbibliography

@misc{boogen2021estimating,
  title={Estimating residential electricity demand: New empirical evidence},
  author={Boogen, N. and Datta, S. and Filippini, M.},
  journal={Energy Policy},
 
  year={2021}
}

@misc{meier1999repraesentative,
  title={Repraesentative VDEW-Lastprofile},
  author={Meier, H. and Fünfgeld, C. and Adam, T. and Schieferdecker, B.},
  year={1999}
}

@misc{bruckmeier2017teilbericht,
  title={Teilbericht Basisdaten Projekt MONA 2030: Grundlage für die Bewertung von Netzoptimierenden Maßnahmen},
  author={Bruckmeier, A. and Böing, F. and Hinterstocker, M. and Kleinertz, B. and Konetschny, C. and Müller, M. and Samweber, F.},
  year={2017}
}

@misc{bdew_stromverbrauch,
  title={Stromverbrauch in Deutschland},
  author={BDEW},
  note={Available: \url{https://www.bdew.de/service/daten-und-grafiken/stromverbrauch-deutschland/}}
}

@misc{scholz2017how,
  title={How to improve Standard Load Profiles: Updating, Regionalization and Smart Meter Data},
  author={Scholz, D. and Müsgens, F.},
  booktitle={14th International Conference on the European Energy Market (EEM)},
  year={2017}
}

@misc{hinterstocker2014bewertung,
  title={Bewertung der aktuellen Standardlastprofile Österreichs und Analyse zukünftiger Anpassungsmöglichkeiten im Strommarkt},
  author={Hinterstocker, M. and Rau, M. and von Roon, S.},
  booktitle={Symposium Energieinnovation},
  year={2014}
}

@misc{staats2018optimierung,
  title={Optimierung der Fahrplanprognose für die Beschaffung elektrischer Energie durch Einsatz messtechnisch ermittelter Kundenlastgänge anstelle von Standardlastprofilen},
  author={Staats, J. and Watts, D. and Bruce-Boye, C.},
  year={2018}
}

@misc{sinoquet2021time,
  title={Time Series Analysis and Modeling to Forecast: a Survey},
  author={Sinoquet, C. and Dama, F.},
  journal={CoRR},
  year={2021}
}

@book{taylor2017forecasting,
  title={Forecasting at Scale},
  author={Taylor, S. J. and Letham, B.},
  year={2017}
}

@misc{logarithmo_openmeter,
  title={OpenMeter},
  author={logarithmo GmbH \& Co.KG},
  note={Available: \url{www.openmeter.de}}
}

@misc{gabrielski2023markov,
  title={A Markov Chain Model for Imputation of Electricity Consumption Time Series},
  author={Gabrielski, J. and Häger, U.},
  booktitle={2023 58th International Universities Power Engineering Conference (UPEC)},
  year={2023}
}

@misc{baranek2013optimierung,
  title={Optimierung der Lastprognose mittels Smart Meter Daten},
  author={Baranek, D. and Probst, A. and Tenboh, S.},
  
  year={2013}
}

@misc{pfahl2024erwerbstaetigenquoten,
  title={Erwerbstätigenquoten und Erwerbsquoten 1991–2022},
  author={Pfahl, S. and Unrau, E.},
 
  year={2024}
}

@misc{rousseeuw1987silhouettes,
  title={Silhouettes: A graphical aid to the interpretation and validation of cluster analysis},
  author={Rousseeuw, P. J.},
  journal={Journal of Computational and Applied Mathematics},

  year={1987}
}

@misc{savitzky1964smoothing,
  title={Smoothing and Differentiation of Data by Simplified Least Squares Procedures},
  author={Savitzky, A. and Golay, M.},
  journal={Analytical Chemistry},
  year={1964}
}

@misc{odyssee_mure,
  author    = {{Odyssee Database}},
  year      = {2024},
  howpublished = {\url{https://www.odyssee-mure.eu/publications/efficiency-by-sector/households/}},
}
\end{document}